# Charged ferroelectric domain walls for deterministic a.c. signal control

*J. Schultheiß[1, *], E. Lysne[1], L. Puntigam[2], J. Schaab[3], E. Bourret[4], Z. Yan[4,5], S. Krohns[2], and D. Meier[1, *]*

[1] Department of Materials Science and Engineering, Norwegian University of Science and Technology (NTNU), 7034, Trondheim, Norway
[2] Experimental Physics V, University of Augsburg, 86159, Augsburg, Germany
[3] Department of Materials, ETH Zurich, 8093, Zurich, Switzerland
[4] Materials Science Division, Lawrence Berkeley National Laboratory, 94720, Berkeley, CA, USA
[5] Department of Physics, ETH Zurich, 8093, Zurich, Switzerland
*corresponding authors: jan.schultheiss@ntnu.no; dennis.meier@ntnu.no

**The direct current (d.c.) conductivity and emergent functionalities at ferroelectric domain walls are closely linked to the local polarization charges. Depending on the charge state, the walls can exhibit unusual d.c. conduction ranging from insulating to metallic-like, which is leveraged in domain-wall-based memory, multi-level data storage, and synaptic devices. In contrast to the functional d.c. behaviors at charged walls, their response to alternating currents (a.c.) remains to be resolved. Here, we reveal a.c. characteristics at positively and negatively charged walls in ErMnO$_3$, distinctly different from the response of the surrounding domains. By combining voltage-dependent spectroscopic measurements on macroscopic and local scales, we demonstrate a pronounced non-linear response at the electrode-wall junction, which correlates with the domain-wall charge state. The dependence on the a.c. drive voltage enables reversible switching between uni- and bipolar output signals, providing conceptually new opportunities for the application of charged walls as functional nano-elements in a.c. circuitry.**

Ferroelectric domain walls are excellent candidates for the development of next-generation nanoelectronics, exhibiting a thickness that approaches the unit cell level.[1-3] Similar to 2D systems such as graphene,[4] MoS$_2$ single layers,[5] and the LaAlO$_3$/SrTiO$_3$ hetero-interface,[6] they display unique electronic transport properties[3] and large carrier mobilities.[7] In addition to their transport properties, the ferroelectric domain walls are spatially mobile and can be injected and deleted on demand, which enables them to take an active role as reconfigurable elements in, e.g., memory,[8,9] diode[10] or memristor[11] devices. Recently, it was demonstrated that intrinsic electronic correlation phenomena at ferroelectric domain walls can be used to control electrical currents, removing the need to write and erase the walls.[12,13] This observation promoted the idea to develop the walls themselves into devices instead of using them as active elements in much larger electronic components. The approach is intriguing as it breaks the mold of classical device architectures, taking full advantage of the ultra-small feature size of ferroelectric domain walls. Compared to more than a decade of research on domain-wall devices that operate based on the injection and deletion of domain walls,[14,15] however, little is known about the technological potential of stationary walls. Only recently, it was shown that ferroelectric domain walls can be used to emulate the behavior of electronic components at the nanoscale, acting as binary switches[12] and half-wave rectifiers[13]. First insight into the electronic properties of domain walls under alternating currents (a.c.) was obtained for neutral domain walls in the gigahertz regime[16-20] and applications as tunable microwave devices and acoustic wave filters have been suggested.[21] In contrast, charged domain walls, which exhibit unusual conduction properties under direct current (d.c.), have been found to be electronically inactive at high frequencies in the gigahertz regime.[16,22]

In this work, we study the electronic a.c. response at positively and negatively charged ferroelectric domain walls at intermediate frequencies in the kilo- and megahertz regime. Performing nanoscale spectroscopic measurements on ErMnO$_3$, we observe domain-wall specific cut-off frequencies, $f_c$, at which the current-voltage characteristic of the electrode-wall junction changes from asymmetric to symmetric. By varying the a.c. voltage amplitude applied to negatively charged walls, we show that the cut-off frequency can readily be tuned by about one order of magnitude. This tunability enables reversible switching between uni- and bipolar output signals, facilitating active signal conversion in a.c. circuits at the nanoscale.

## a.c. response of positively and negatively charged walls

Hexagonal ErMnO$_3$ is a ferroelectric small band gap semiconductor (p-type, $E_{gap} \approx 1.6$ eV).[23-25] The spontaneous polarization is parallel to the c-axis ($P \approx 6$ µC/cm$^2$)[26] and originates from a structural[27,28] lattice-trimerization, leading to explicitly robust ferroelectric domain walls, including all fundamental types of 180° walls (i.e., neutral side-by-side walls, positively charged head-to-head walls, and negatively charged tail-to-tail walls).[29] The conduction of the neutral walls has been intensively investigated both in the d.c.[29-31] and a.c.[13,16] regimes continuously covering frequencies up to the gigahertz range and their basic electronic properties are well understood. In contrast, at charged domain walls only the d.c. transport behavior[29,31,32] and the response at high frequencies in the microwave range[16] have been studied, whereas their a.c. properties at intermediate frequencies remain to be explored.

The electrical d.c. transport of a (110)-oriented ErMnO$_3$ single crystal (in-plane polarization) is displayed in the conductive atomic force microscopy (cAFM) map in Figure 1a. The orientation of the ferroelectric polarization is indicated by the arrows, determined from the calibrated piezoresponse force microscopy (PFM) image displayed in the inset of Figure 1a. The data shows the established transport behavior,[29] that is, enhanced conductance (bright) at the tail-to-tail walls and reduced conductance (dark) at the head-to-head walls. To investigate the electronic properties of the charged domain walls in the kilo- to megahertz regime, we perform AC-cAFM[13] scans at the same position. AC-cAFM measures the d.c. response ($I_{d.c.}^{out}$) under applied bipolar voltages ($V_{a.c.}^{in}$) as function of frequency (method section and Figure S1). $V_{a.c.}^{in}$ describes the amplitude of the bipolar voltage. Figure 1b presents the characteristic AC-cAFM response of both head-to-head and tail-to-tail domain walls at a frequency $f = 0.5$ MHz. In contrast to previous measurements performed under microwave frequencies,[16] a pronounced response to the a.c. voltage is detected at the charged domain walls, clearly separating them from the surrounding domains. In addition, the scan in Figure 1b reveals a significant difference in the AC-cAFM response at walls with opposite charge state, showing reduced and enhanced current signals at the head-to-head and tail-to-tail walls, respectively. Thus, the behavior observed in the kilohertz-frequency AC-cAFM scan is consistent with the d.c. current distribution probed by cAFM (Figure 1a) which is expected to be approached for $f \to 0$ Hz.

A systematic analysis of $I_{d.c.}^{out}$ at charged domain walls as a function of the frequency of the applied a.c. voltage is presented in Figures 1c and d. Figure 1c displays $I_{\text{d.c.}}^{\text{out}}$ on a logarithmic frequency scale recorded along the solid line indicated in Figure 1b, featuring a direct comparison of tail-to-tail and head-to-head domain walls with respect to the surrounding domains. At $f = 1$ MHz, $I_{\text{d.c.}}^{\text{out}}$ at the insulating head-to-head domain wall is suppressed in comparison to the domains, whereas an enhancement of $I_{\text{d.c.}}^{\text{out}}$ is observed at the tail-to-tail domain wall. With increasing frequency, $I_{\text{d.c.}}^{\text{out}}$ reveals a step-like feature indicating a relaxation-like process (Figure 1d).[13] As indicated by the smaller arrows, a cut-off frequency $f_c$ is defined above which $I_{\text{d.c.}}^{\text{out}}$ vanishes. The cut-off frequency $f_c$ marks a qualitative change in the current-voltage characteristics. Analogous to previous measurement at neutral domain walls in ErMnO$_3$,[13] the a.c. response at $f < f_c$ is asymmetric due to the Schottky-like tip-sample contact, leading to a non-zero current signal in AC-cAFM.[33,34] For $f > f_c$, the AC-cAFM contrast vanishes, indicating symmetric current-voltage characteristics. Furthermore, for the conductive tail-to-tail domain wall the cut-off frequency ($f_c^{\leftrightarrow} \sim 4.0$ MHz) is about four times higher than for the domains ($f_c^{\text{Domain}} \sim 1.0$ MHz). Consistent with its reduced d.c. conductance (Figure 1a), the cut-off frequency of the insulating head-to-head domain wall is below the lower limit of the frequency range of the measurement ($f_c^{\rightarrow\leftarrow} < 0.1$ MHz). Thus, we focus on the tail-to-tail walls in the following in-depth analysis.

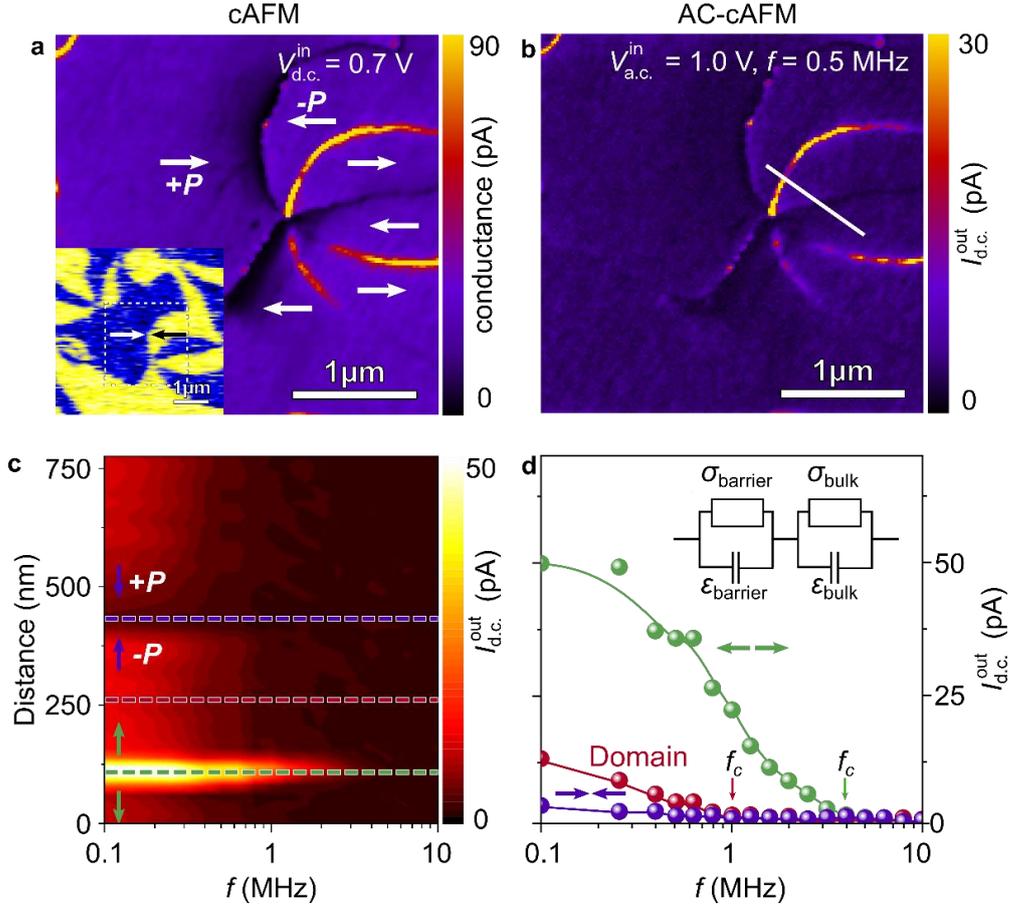

**Figure 1 | a.c. response of charged ferroelectric domain walls in ErMnO$_3$. a,** cAFM image displaying reduced and enhanced d.c. conductance at head-to-head and tail-to-tail domain walls, respectively. The polarization direction (indicated by the arrows) is obtained from calibrated PFM data, provided in the inset (blue: +$P$; yellow: -$P$). **b,** AC-cAFM scan taken at the same position as the cAFM image in **a**. **c,** Frequency-dependent evolution of the AC-cAFM signal along the solid line in **b**. Pronounced AC-cAFM contrast is observed at $f = 0.1$ MHz, vanishing towards increasing frequencies. **d,** Local frequency dependent AC-cAFM response evaluated along the dashed lines in **c** for a domain, a head-to-head, and tail-to-tail domain wall, indicating different cut-off frequencies, $f_c$, (indicated by arrows) above which the respective signals disappear ($f_c^{\leftrightarrow} > f_c^{\text{Domain}} > f_c^{\rightarrow\leftarrow}$). The equivalent circuit model in the inset allows for relating the frequency drop to the local intrinsic conductivity,[13] i.e., $\sigma_{\text{bulk}}^{\leftrightarrow} > \sigma_{\text{bulk}}^{\text{Domain}} > \sigma_{\text{bulk}}^{\rightarrow\leftarrow}$ (barrier conductivity, $\sigma_{\text{barrier}}$; barrier permittivity, $\varepsilon_{\text{barrier}}$; bulk conductivity, $\sigma_{\text{bulk}}$; and bulk permittivity, $\varepsilon_{\text{bulk}}$).

To rationalize the behavior probed at the charged domain walls, we apply the same equivalent circuit model as used in ref. 13 (see inset to Figure 1d). Here, two *RC* elements are connected in series. The domains and domain walls are described by a resistor (with conductivity $\sigma_{\text{bulk}}$) in parallel with a capacitor (with permittivity $\varepsilon_{\text{bulk}}$). The probed volume within the bulk can be approximated by a hemisphere with radius of about 600 nm.[35] The barrier between tip and sample is described by a barrier conductivity ($\sigma_{\text{barrier}}$) connected in parallel with a capacitor (with permittivity $\varepsilon_{\text{barrier}}$).[26,36,37] For $f < f_c$, the transport behavior is dominated by the diode-like tip-sample contact, leading to asymmetric current-voltage characteristics and, hence, a pronounced current signal $I_{\text{d.c.}}^{\text{out}}$ in AC-cAFM. For higher frequencies ($f > f_c$), the $I_{\text{d.c.}}^{\text{out}}$ contrast vanishes, indicating that the tip-sample contact gets short-circuited via the barrier capacitance.[38] Within this model,[13,36,39] the cut-off frequency is given by the relation

$$f_c \propto \frac{\sigma_{\text{bulk}}}{2\pi\epsilon_0\epsilon_{\text{barrier}}}, \qquad (1)$$

connecting the measured cut-off frequencies and the intrinsic electronic transport properties. Note that relation (1) is an approximation that is valid for $\sigma_{\text{bulk}} \gg \sigma_{\text{barrier}}$ and $\varepsilon_{\text{barrier}} \gg \varepsilon_{\text{bulk}}$; in other cases, additional contributions are to be considered as explained elsewhere[36,40]. Thus, the AC-cAFM data in Figure 1b-d corroborate the intrinsic nature of the enhanced (reduced) conductivity at tail-to-tail (head-to-head) walls ($f_c^{\leftrightarrow} > f_c^{\text{Domain}} > f_c^{\rightarrow\leftarrow} \rightarrow \sigma_{\text{bulk}}^{\leftrightarrow} > \sigma_{\text{bulk}}^{\text{Domain}} > \sigma_{\text{bulk}}^{\rightarrow\leftarrow}$).

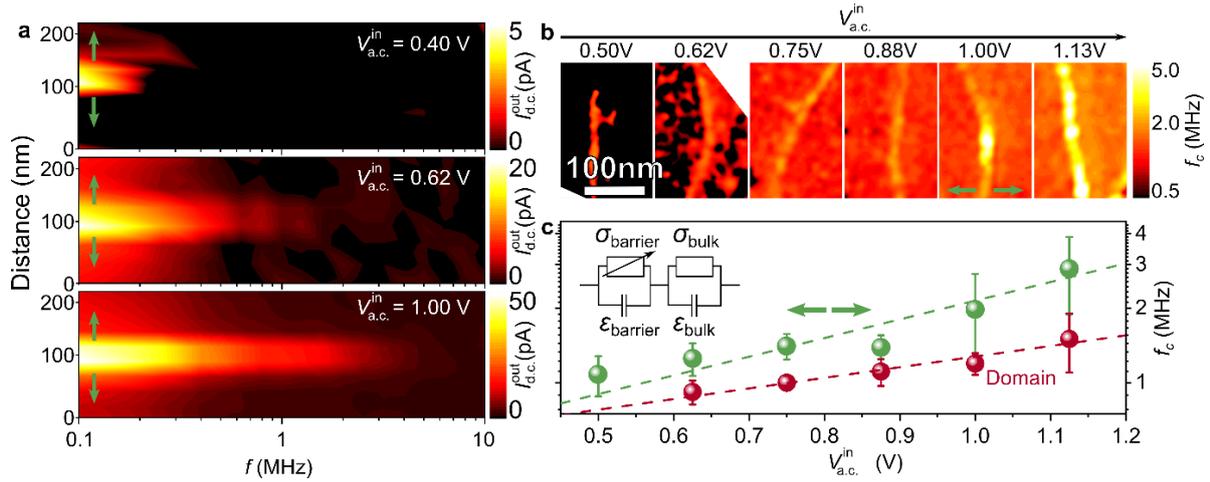

**Figure 2 | Relation between drive voltage and cut-off frequency. a,** Cross-sectional data showing the frequency dependence of the AC-cAFM response at a negatively charged tail-to-tail domain wall for different voltages. With increasing voltage, $f_c$ shifts to higher frequencies. **b,** Spatially resolved measurements of $f_c$, recorded at different tail-to-tail domain walls (see also Figure S2). The data is derived from a series of AC-cAFM images with logarithmically increasing frequencies by fitting the current decay pixel by pixel as explained in the main text and Figure S3. **c,** Comparison of the voltage dependence of the cut-off frequencies measured at tail-to-tail walls and in the surrounding domains. Plotted are the mean values; error bars represent the standard deviation. The dashed lines display a guide to the eye. A nonlinear barrier conductivity is introduced into the equivalent circuit model, as displayed in the inset, which allows for capturing the observed voltage-dependent behavior.[42]

**Voltage-dependent a.c. response at tail-to-tail domain walls**

The effect of varying drive voltage on the cut-off frequency $f_c$ is presented in Figure 2, showing an overview of frequency- and voltage-dependent AC-cAFM measurements for conducting tail-to-tail domain walls (see Figure S2 for complementary cAFM and PFM data). Figure 2a displays spatially resolved data measured along tail-to-tail domain walls with different $V_{\text{a.c.}}^{\text{in}}$. To avoid possible artefacts caused by repeatedly scanning the same area,[41] the measurements are performed at different positions on selected walls with comparable d.c. conductance (see Figure S2 for details). We observe that $f_c$ increases with increasing $V_{\text{a.c.}}^{\text{in}}$, shifting by more than one order of magnitude as $V_{\text{a.c.}}^{\text{in}}$ is raised from 0.40 V to 1.00 V.

To systematically analyze the correlation between $f_c$ and $V_{\text{a.c.}}^{\text{in}}$, we record frequency-dependent AC-cAFM maps for a wider voltage range from which we calculate $f_c$ pixel by pixel as explained in the Supplementary Information (see Figure S3). Figure 2b displays the resulting cut-off-frequency maps for six tail-to-tail domain walls and the surrounding domains measured at different $V_{\text{a.c.}}^{\text{in}}$. The mean cut-off frequencies obtained for the domains and domain walls are displayed in Figure 2c, increasing as function of the applied voltage. Importantly, the pronounced voltage dependence of the cut-off frequency cannot be reproduced based on Equation (1), which indicates an additional contribution.[40]

To clarify the origin of the additional drive-voltage dependence revealed by AC-cAFM, we perform complementary voltage-dependent macroscopic spectroscopy experiments. The loss factor, $\tan\delta$, from $10^{-4}$ to 2 MHz is shown in Figure 3. The voltage and frequency dependence of the dielectric permittivity and the conductivity is displayed in Figure S4. For all applied voltages, we observe two characteristic regimes, which relate to the electrode-sample interface ($f \lesssim 5 \cdot 10^{-2}$ MHz) and the internal properties of ErMnO$_3$ ($f \gtrsim 5 \cdot 10^{-2}$ MHz) as demonstrated elsewhere.[37,43] We find that the frequency response above $5 \cdot 10^{-2}$ MHz is unaffected by the applied voltage, whereas the low-frequency response ($f \lesssim 5 \cdot 10^{-2}$ MHz) displays a pronounced voltage dependence. Analogous to the local measurements (Figure 1 and 2), we define a cut-off frequency $f_c$ ($\tan\delta$ falls below 25 % of the maximum value,[13] Figure 3b) which takes the broadness of the relaxation peak into account.[44] The value $f_c$ represents a measure for the frequency at which the barrier contributions are short-circuited. In the macroscopic measurements, we find a voltage-independent cut-off frequency $f_c = 1.3$ MHz, which is consistent with the cut-off frequencies identified for the domains in the local AC-cAFM measurements. As indicated by the solid lines in Figure 3 and Figure S4, the macroscopic dielectric response can be described via fits using the equivalent circuit model displayed in the inset of Figure 2c (see Method section). The analysis shows that $\sigma_{barrier}$ increases by more than one order of magnitude when $V_{a.c.}^{in}$ is increased from 1 V to 20 V, while all other parameters remain almost unchanged.

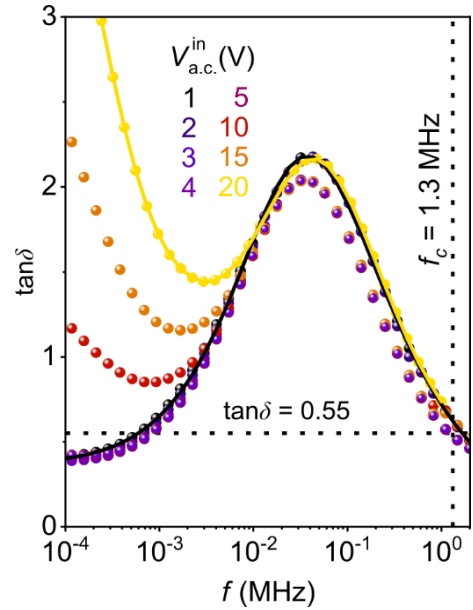

**Figure 3 | Voltage- and frequency dependent macroscopic dielectric response.** Voltage dependence of the loss factor, $\tan\delta$, as a function of frequency, gained on the same sample used for the local measurements in Figures 1 and 2. The solid lines represent fits of the experimental data (for $V_{\text{a.c.}}^{\text{in}} = 1$ V and $V_{\text{a.c.}}^{\text{in}} = 20$ V) utilizing the equivalent circuit model displayed in Figure 1c extended by a frequency-dependent resistance for the bulk as explained previously (see method section).[26,37] Analogous to the local measurements (Figures 1 and 2), we define a cut-off frequency, $f_c$, at which the barrier is short-circuited and the bulk response dominates ($\tan\delta$ falls below 25 % of the maximum value). The identified value of $f_c = 1.3$ MHz is in agreement with the cut-off frequency of the domains found in AC-cAFM (Figure 2c).

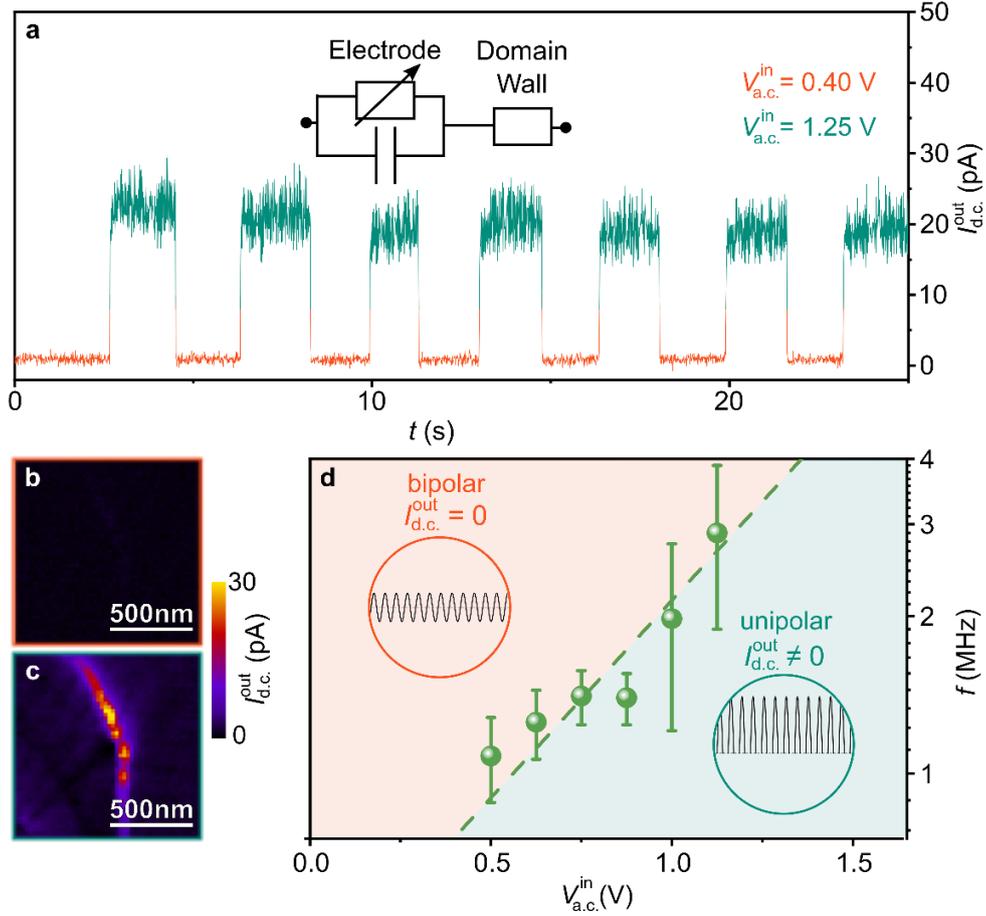

**Figure 4 | Reversible control of the a.c. response at tail-to-tail domain walls. a,** AC-cAFM current signal measured with a stationary tip placed on a tail-to-tail domain wall as function of time over multiple cycles, switching between bipolar (symmetric, $V_{a.c.}^{in} = 0.40$ V, $I_{d.c.}^{out} = 0$) and unipolar (asymmetric, $V_{a.c.}^{in} = 1.25$ V, $I_{d.c.}^{out} \neq 0$) response. A schematic illustration of a two-terminal a.c. element emulated by the electrode-wall junction with its equivalent circuit representation[42] is displayed in the inset. The spatially resolved AC-cAFM image gained for the bipolar and unipolar output of the a.c. element is displayed in **b** and **c**, respectively. **d,** Summary of the electronic response of the a.c. element in relation to the cut-off frequency and the applied bipolar voltage. The data points and error bars represent $f_c$ (taken from Figure 2c) and mark the transition between a bipolar ($I_{d.c.}^{out} = 0$) and unipolar ($I_{d.c.}^{out} \neq 0$) output signal. This transition between the two distinctly different regimes can either be driven by a change in $V_{a.c.}^{in}$ ($f$ = const.) or, vice versa, by changing $f$ ($V_{a.c.}^{in}$ = const.).

This leads us to the conclusion that the voltage-dependent AC-cAFM response in Figure 2 originates from the Schottky-like nature of the tip-sample contact. The latter is corroborated by the equivalent circuit fitting of the macroscopic dielectric data (Figure S2), which indicates a substantial voltage-driven barrier lowering (Figure S4).[45,46] Thus, the AC-cAFM data expands previous macroscopic studies on dielectrics[47,48] to the nanoscale. The voltage dependence of $f_c$ (Figure 2c) is captured by considering the non-linear voltage dependence of the barrier conductivity in the equivalent circuit model sketched in the inset to Figure 2c, which leads to[36]

$$f_c(V_{a.c.}^{in}) \propto \frac{\sigma_{bulk} + \sigma_{barrier}(V_{a.c.}^{in})}{2\pi\epsilon_0\epsilon_{barrier}}, \quad (2)$$

showing that $f_c$ shifts to higher frequencies with increasing voltage. In summary, our studies show that the a.c. characteristics observed at the tail-to-tail domain walls result from their enhanced intrinsic conductivity (Figure 1) in combination with the formation of a voltage-dependent barrier at the electrode–wall junction (Figure 2 and 3). The shift of $f_c$ with $V_{a.c.}^{in}$ becomes observable in the local AC-cAFM measurements due to a higher local electric field ($E \approx 40$ kV/cm) compared to the electric fields ($E \approx 0.4$ kV/cm) used in the macroscopic measurements.

**Reversible voltage-driven control of the a.c. response**

The relation between $V_{a.c.}^{in}$ and the response at the tail-to-tail domain wall allows for controlling the local electronic transport characteristics. In Figure 4, we demonstrate how the junction between the electrode and the ferroelectric domain wall can be utilized to reversibly switch between uni- and bipolar output signals. The AC-cAFM data in Figure 4a is recorded at constant frequency ($f = 1$ MHz) as function of time, varying $V_{a.c.}^{in}$ repeatedly between 0.40 V (orange) and 1.25 V (green) while keeping the probe tip stationary at the position of the wall. Depending on the applied voltage amplitude, we measure two qualitatively different responses, switching between symmetric ($I_{d.c.}^{out} = 0$) and asymmetric ($I_{d.c.}^{out} \neq 0$) responses. Spatially resolved AC-cAFM scans obtained at the same tail-to-tail domain wall at $V_{a.c.}^{in} = 0.4$ V and $V_{a.c.}^{in} = 1.25$ V are displayed as insets to Figure 4a, showing the same switching behavior consistent with the data presented in Figure 2a.

The two-terminal a.c. element emulated by the electrode–wall junction and the respective equivalent circuit model are sketched in Figures 4b and c. The electrode-wall junction responds symmetrically at low $V_{a.c.}^{in}$, whereas an asymmetric response is detected for high $V_{a.c.}^{in}$. Because of the distinct a.c. response of the electrode–wall junction, the change in voltage allows reversible switching between unipolar and bipolar output. The dependence of $I_{d.c.}^{out}$ on both the applied voltage amplitude and frequency is summarized in Figure 4d. The graph emphasizes the existence of two regimes where the electrode–wall junction exhibits qualitatively different electronic responses. The voltage required to transit between these two regimes can be tuned via the frequency of the input signal. Vice versa, the cut-off frequency can be selected by adjusting the voltage amplitude of the input signal.

The electronic tunability of the diode-like properties at the tail-to-tail walls represents an additional degree of freedom, enabling the design of domain-wall based a.c. electronic components with ultra-small feature size. Application opportunities range from domain-wall based thyrectors that can buffer ripple currents and diodes in transponder circuitry to walls acting as the interconnect between active and passive devices in a.c. nanoelectronics. In general, the application of charged domain walls in low-frequency nanoelectronics offers several advantages compared to their neutral counterparts.[13] In contrast to the neutral walls, which owe their transport properties to the accumulation and/or depletion of ionic defects,[13,30] the conduction at charged domain walls is driven by bound polarization charges, that is, an intrinsic mechanism. The latter implies that defect migration and effects from mixed ionic-electric condictivity[49] play a less important role compared to neutral domain walls, which is important in order to ensure a reversible and deterministic electronic response at the electrode–wall junction. Furthermore, the bound polarization charges can be used as quasi-dopants[50] to tune the local conductivity and, thereby, engineering the electronic properties of the electrode–wall junction on demand. Our work introduces charged ferroelectric domain walls as versatile building blocks for a.c. nanoelectronics in the kilo- to megahertz regime, establishing innovative concepts for domain-wall based nanotechnology and the downscaling of electronic a.c. components in general.

**Acknowledgements**

J.S. acknowledges the support of the Alexander von Humboldt Foundation through the Feodor-Lynen fellowship. D.M. thanks NTNU for support through the Onsager Fellowship Program, the Outstanding Academic Fellow Program, and acknowledges funding from the European Research Council (ERC) under the European Union's Horizon 2020 Research and Innovation Program (Grant Agreement No. 86691). L.P. and S.K. acknowledge funding of the German Science foundation via the Collaborative Research Center TRR80.

**Author contributions**

J.S. and E.L. recorded the scanning probe microscopy data supervised by D.M.. L.P. performed the macroscopic dielectric spectroscopy measurements under supervision of S.K.. E.B. and Z.Y. provided the sample. J.S. and D.M. interpreted the data, devised the device concept, and wrote the manuscript. All authors discussed the results and contributed to the final version of the manuscript.

**Methods**

**Material.** High-quality single crystals were grown using the pressurized floating-zone method[51] and oriented by Laue diffraction with the polarization vector in-plane. The samples with a thickness of about 1 mm were cut and lapped with a 9 µm-grained $Al_2O_3$ water suspension and polished using silica slurry (Ultra-Sol® 2EX, Eminess Technologies, Scottsdale, AZ, USA) to produce a flat surface with a mean roughness of about 1.55 nm (determined by atomic force microscopy considering a 25x25 µm² scan area).

**Local electric characterization.** Scanning probe microscopy measurements were performed on an NT-MDT Ntegra Prisma system (NT-MDT, Moscow, Russia). Voltages (a.c. and d.c.) for PFM, cAFM, and AC-cAFM measurements were applied through the bottom electrode using a function generator (Agilent 33220 A, Santa Clara, CA, USA). All scans were performed using an electrically conductive diamond coated tip (DDESP-10, Bruker, Billerica, MA, USA) with a tip height of 10-15 µm and a maximum tip radius of ≈150 nm. All measurements were carried out at room temperature ($T \approx 25$ °C).

For PFM measurements, the sample was excited using an a.c. voltage ($f = 40$ kHz, $V_{a.c.}^{in} = 1.5$ V), while the laser deflection was read out by lock-in amplifiers (SR830, Stanford Research Systems, Sunnyvale, CA, USA). The PFM response was calibrated on a periodically out-of-plane poled $LiNbO_3$ sample. For cAFM scans a d.c. voltage was applied ($V_{d.c.}^{in} = 0.7$ V), while the sample response was read out using a low current head (SF005, NT-MDT, Moscow, Russia). In this frame, AC-cAFM is an extension of conventional cAFM measurements.[13] Here an a.c. voltage (0.1 MHz < $f$ <10 MHz, 0.4 V < $V_{a.c.}^{in}$ < 1.13 V) is applied to the bottom electrode. The low current head mimics a low pass filter with a cutoff frequency around 1 Hz, making the 0 Hz (also termed d.c. or rectified) component accessible. The recorded analog signal is then transferred to a digital signal and spatial resolution of the 0 Hz component (referred to as $I_{d.c.}^{out}$) is provided by the atomic force microscope (a detailed description is provided in Figure S1).

**Macroscopic dielectric spectroscopy.** Macroscopic dielectric spectroscopy was carried out on the same single crystal also used for local electric characterization. Measurements were performed in a plate capacitor geometry, while both surfaces were coated with silver paint. Measurements were performed using an Alpha Analyzer (Novocontrol, Montabaur, Germany) together with a voltage booster option (HVB300, Novocontrol, Montabaur, Germany), covering a frequency range of $10^{-4}$ to 1 MHz with varying applied bipolar voltages from $V_{a.c.}^{in} = 1 - 20$ V. The measurement was performed at room temperature ($T$~25°C). More details can be found in ref. 43. The fits of the macroscopic data were done using an equivalent circuit model, consisting of two $RC$ circuits connected in series to describe the behavior of the bulk and the barrier independent of each other over the entire frequency regime. Note that the conductivity of the second $RC$ element of this circuit (the internal contributions) consists of a resistor representing the intrinsic conductivity ($\sigma_{bulk}$ in Figure 1d and 2c). In the macroscopic dielectric measurements, an additional contribution to the conductivity for the universal dielectric response ($\sigma_{UDR}$),[37,43] covering the influence of hopping transport on $\sigma'(f) \propto f^n$ with exponent $n < 1$[36,52] was utilized.

# Supplementary information

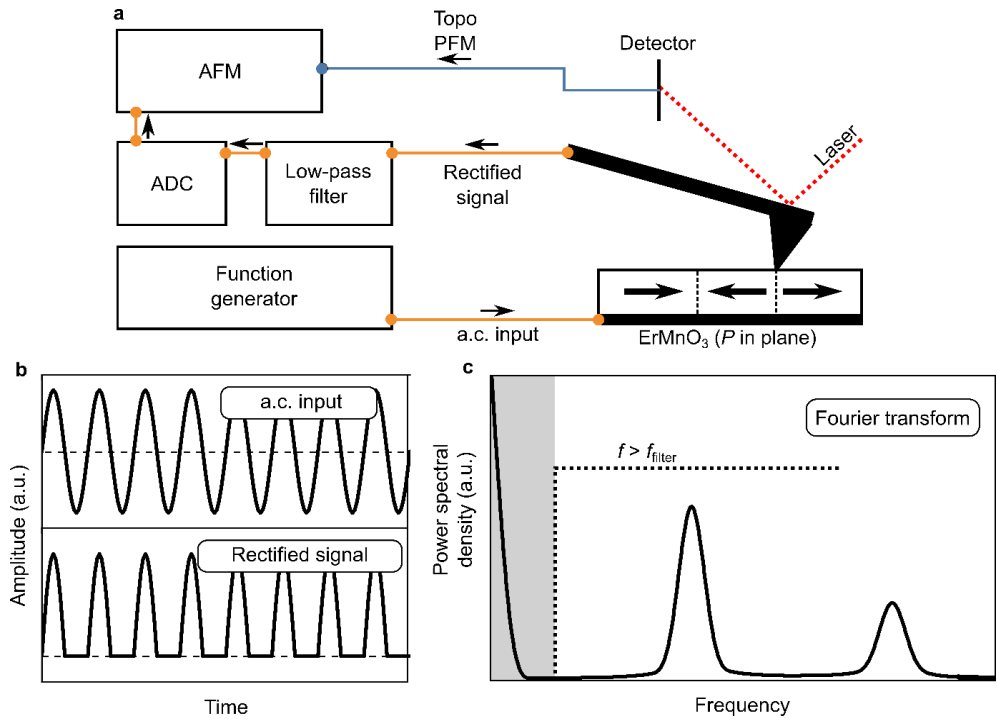

**Figure S1: Principle of the AC-cAFM experiment. a,** Electronic circuit used in AC-cAFM. An a.c. input voltage (with varying frequency, *f*, and amplitude, $V_{a.c.}^{in}$) is applied to the bottom electrode, while a probe tip is scanned over the surface of the sample in contact mode. **b,** Due to the Schottky-like tip-sample contact,[1,2] the alternating input signal gets rectified. **c,** To illustrate the effect of the low-pass filter, the Fourier transform of the rectified signal of Figure (b) is calculated. High-frequency components of the analog signal are filtered out using a low-pass filter with $f_{filter} \sim 1 Hz$. The d.c. component of the signal is then converted into a digital signal using an analog-to-digital (ADC) converter, which is read out by an atomic force microscopy (AFM) unit.

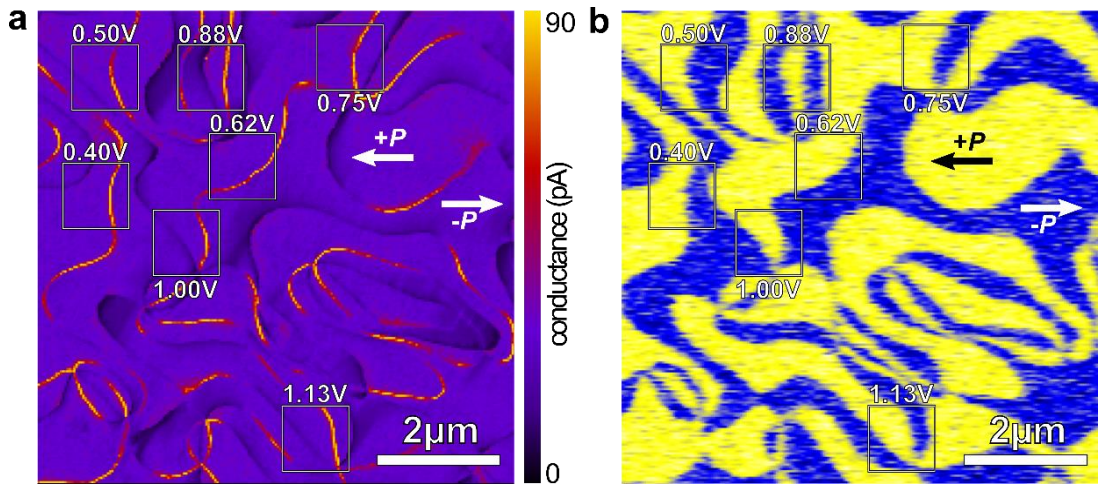

**Figure S2: Positions of the voltage-dependent AC-cAFM scans in Figure 2. a,** cAFM image ($V_{d.c.}^{in} = 0.7$ V) and **b** calibrated PFM overview scan at the same position recorded on a (110)-oriented ErMnO$_3$ single crystal, featuring conductive tail-to-tail and insulating head-to-head domain walls (the polarization orientation, *P*, of the domains is indicated by the arrows). The frequency dependence of the AC-cAFM response of the conductive tail-to-tail domain walls inside the 1.5 x 1.5 µm² boxes was systematically investigated under different $V_{a.c.}^{in}$ (the respective $V_{a.c.}^{in}$ is indicated). The results are displayed in Figure 2. The conductive domain walls were chosen for this comparative study since they have a quantitatively comparable d.c. conductance response.

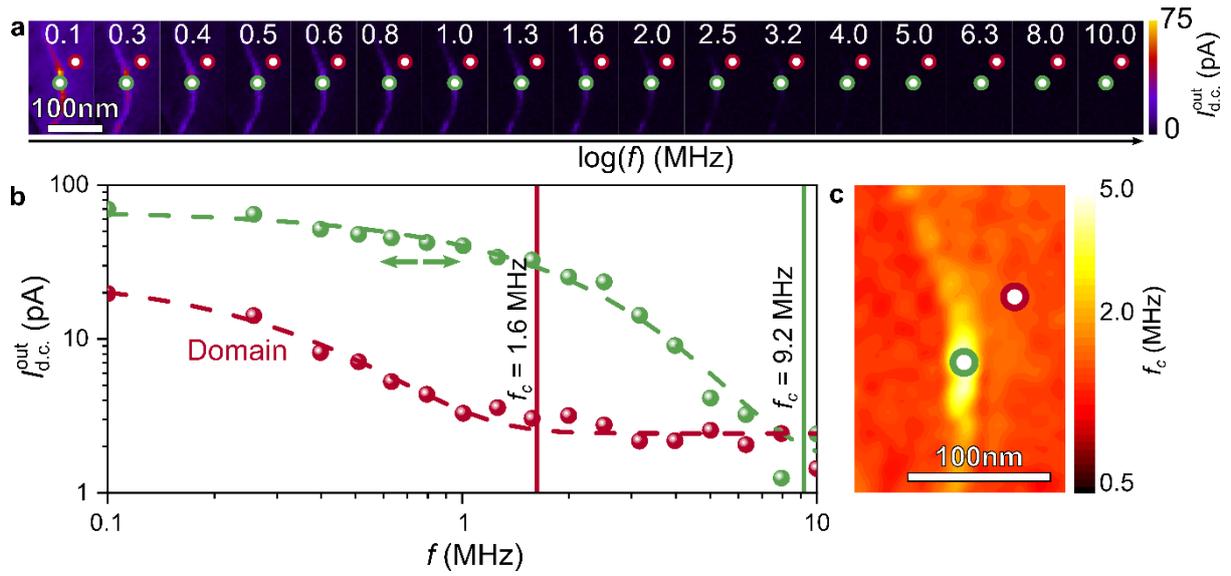

**Figure S3: Spatial resolution of the cut-off frequency. a,** A series of AC-cAFM scans is displayed at logarithmically increasing a.c. frequencies at the red and green marked positions at the tail-to-tail domain wall and within a domain. The frequency values (in MHz) are displayed in the respective AC-cAFM images. **b,** The AC-cAFM contrast is evaluated for these two positions (one pixel corresponding to an area, $A \sim 71 \cdot 10^3$ nm$^2$, estimated by the radius of the tip ($r \approx 150$ nm)). An exponential decay function ($f(f) = a \cdot \exp(-f/f_0) + c$) is fitted to the experimental data, with $a$, $f_0$, and $c$ as fitting parameters. Here, $f_0$ represents the frequency at which 37% of the initial value of the AC-cAFM contrast is reached.[3] Exemplary fits are displayed in **b** as dashed lines. The cut-off frequency, at which the AC-cAFM contrast vanishes, is finally calculated as $f_c = 5 \cdot f_0$ and is displayed for the two pixels as solid vertical lines in **b**. The factor 5 was chosen here, because the AC-cAFM signal reaches a value of less than 1 % of its original value.[3] The analysis was automatized for each pixel within the chosen area using a MatLab program to display the cut-off frequency spatially resolved, exemplarily shown for the investigated scan area in **c**. Figure 2b displays the results for a broad range of voltages.

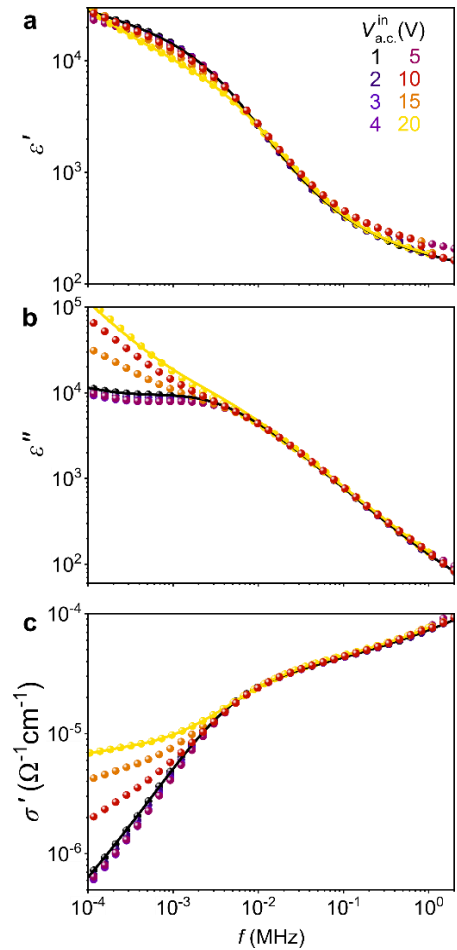

**Figure S4: Frequency- and voltage-dependence of the macroscopic permittivity and conductivity.** The frequency dependent permittivity (real, $\varepsilon'$, and imaginary, $\varepsilon''$, part) and conductivity, $\sigma'$, are displayed in **a, b,** and **c,** respectively. The data is measured under different voltages, $V_{\text{a.c.}}^{\text{in}}$. The solid lines represent

fits of the experimental data utilizing the equivalent circuit model displayed in the inset of Figure 2c (extended by the universal dielectric response,[4,5] as described in the methods section).